\documentclass[12pt]{article}%
\usepackage{url}
\usepackage{amsmath}
\usepackage{graphicx}
\usepackage{amsfonts}
\usepackage{amssymb}%
\begin{document}

\title{Cardiac Alternans Arising from an Unfolded Border-Collision Bifurcation}
\author{Xiaopeng Zhao$^{1}$\thanks{Now at
Mechanical, Aerospace, and Biomedical Engineering Department,
University of Tennessee. Email: xzhao9@utk.edu.}, David G.
Schaeffer$^{2}$, Carolyn M. Berger$^{3}$,\\Wanda Krassowska$^{1}$,
and Daniel J. Gauthier$^{1,3}$\\Department of Biomedical
Engineering$^{1}$, Mathematics$^{2}$, and Physics$^{3}$\\and Center
for Nonlinear and Complex Systems\\Duke University, Durham, NC
27708}
\date{}
\maketitle

\begin{abstract}
Following an electrical stimulus, the transmembrane voltage of cardiac tissue
rises rapidly and remains at a constant value before returning to the resting
value, a phenomenon known as an action potential. When the pacing rate of a
periodic train of stimuli is increased above a critical value, the action
potential undergoes a period-doubling bifurcation, where the resulting
alternation of the action potential duration is known as alternans in the
medical literature. Existing cardiac models treat alternans either as a smooth
or as a border-collision bifurcation. However, recent experiments in paced
cardiac tissue reveal that the bifurcation to alternans exhibits hybrid
smooth/nonsmooth behaviors, which can be qualitatively described by a model of
so-called unfolded border-collision bifurcation. In this paper, we obtain
analytical solutions of the unfolded border-collision model and use it to
explore the crossover between smooth and nonsmooth behaviors. Our analysis
shows that the hybrid smooth/nonsmooth behavior is due to large variations in
the system's properties over a small interval of the bifurcation parameter,
providing guidance for the development of future models.

\end{abstract}

\section{Introduction}

\subsection{Background}

Cardiovascular disease is the number one cause of death in the United States
\cite{AHA}. Over half of the mortality is due to sudden cardiac arrest that is
often initiated by ventricular fibrillation, a fatal heart rhythm disorder.
The induction and maintenance of ventricular fibrillation has been connected
to the dynamics of local cardiac electrical properties
\cite{Rosenbaum1994NEJM,Pastore1999Cir}. Therefore, studying cardiac dynamics
is important for understanding life-threatening arrhythmias and developing
therapies for preventing sudden cardiac death.

\begin{figure}[tbh]
\centering
\includegraphics[width=2.5in]{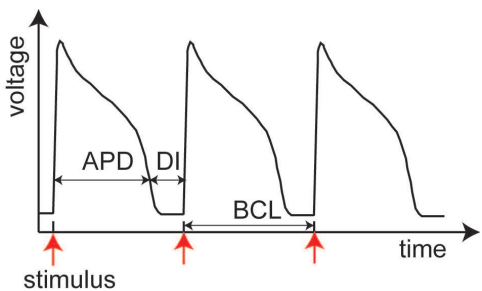}\caption{Schematic action
potential showing the response of the transmembrane voltage to periodic
electrical stimuli.}%
\label{fig:action_potential}%
\end{figure}

To develop an understanding of cardiac rhythm instability, we briefly review
the electrophysiology of the heart. Cardiac cells respond to an electrical
stimulus by eliciting an action potential \cite{plonsey2000}, which consists
of a rapid depolarization of the transmembrane voltage followed by a much
slower repolarization process before returning to the resting value (Fig.
\ref{fig:action_potential}). The time interval during which the voltage is
elevated is called the action potential duration (APD). As shown in Fig.
\ref{fig:action_potential}, the time between the end of an action potential to
the beginning of the next one is called the diastolic interval (DI). The time
interval between two consecutive stimuli is called the basic cycle length
(BCL). 

Under a periodic train of electrical stimuli, the steady-state response
consists of phase-locked action potentials, where each stimulus gives rise to
an identical action potential (1:1 pattern) when the pacing rate is slow. When
the pacing rate becomes sufficiently fast, the 1:1 pattern may be replaced by
a 2:2 pattern, so-called electrical alternans
\cite{gilmour99,garfinkel00,panfilov98}, where the APD\ alternates between
short and long values. Using theory and experiments, a causal connection
between alternans and the vulnerability to fatal cardiac arrhythmias such as
ventricular fibrillation has been established by various authors [2, 3, 5-13].
Therefore, understanding mechanism of alternans is a crucial step in detection
and prevention of fatal arrhythmias.

It has long been hypothesized
\cite{nolasco68,chialvo90,fox02,hall02,tolkacheva04,hall97} that
alternans is mediated by a classical period-doubling bifurcation,
which can be described using a smooth iterated map, and which occurs
when one eigenvalue of the Jacobian crosses the unit circle through
$-1$ \cite{strogatz94,nayfeh95}. We restrict our attention to
supercritical rather than subcritical bifurcations because the
former are observed in most experiments and theoretical models
exhibiting electrical alternans. Based on this hypothesis, various
authors attempted to develop criteria for the onset of alternans
\cite{nolasco68,hall02,fox02} as well as algorithms to control
alternans \cite{hall02,tolkacheva04,hall97}. Recently, a few authors
\cite{sun95,hassouneh04,Berger2005APS,Fenton2006ITP,Cherry2007}
proposed a different hypothesis: alternans may be mediated through a
border-collision period-doubling bifurcation. Border-collision
bifurcations occur in piecewise smooth maps
\cite{diBernardo06,BCbook}. In contrast to classical period-doubling
bifurcations, eigenvalues are not indicative of the onset of a
border-collision period-doubling bifurcation. Instead, a
border-collision bifurcation occurs when a branch of fixed points
collides with a \emph{border}, i.e., a discontinuity surface in
state space. Knowing the mechanism of alternans may help researchers
to choose the proper types of functions to model this instability.
More importantly, to develop model-based control methods requires
knowledge of the underlying dynamics \cite{hassouneh04}.

\begin{figure}[tbh]
\centering
\begin{tabular}
[c]{cc}%
\includegraphics[width=1.5in]{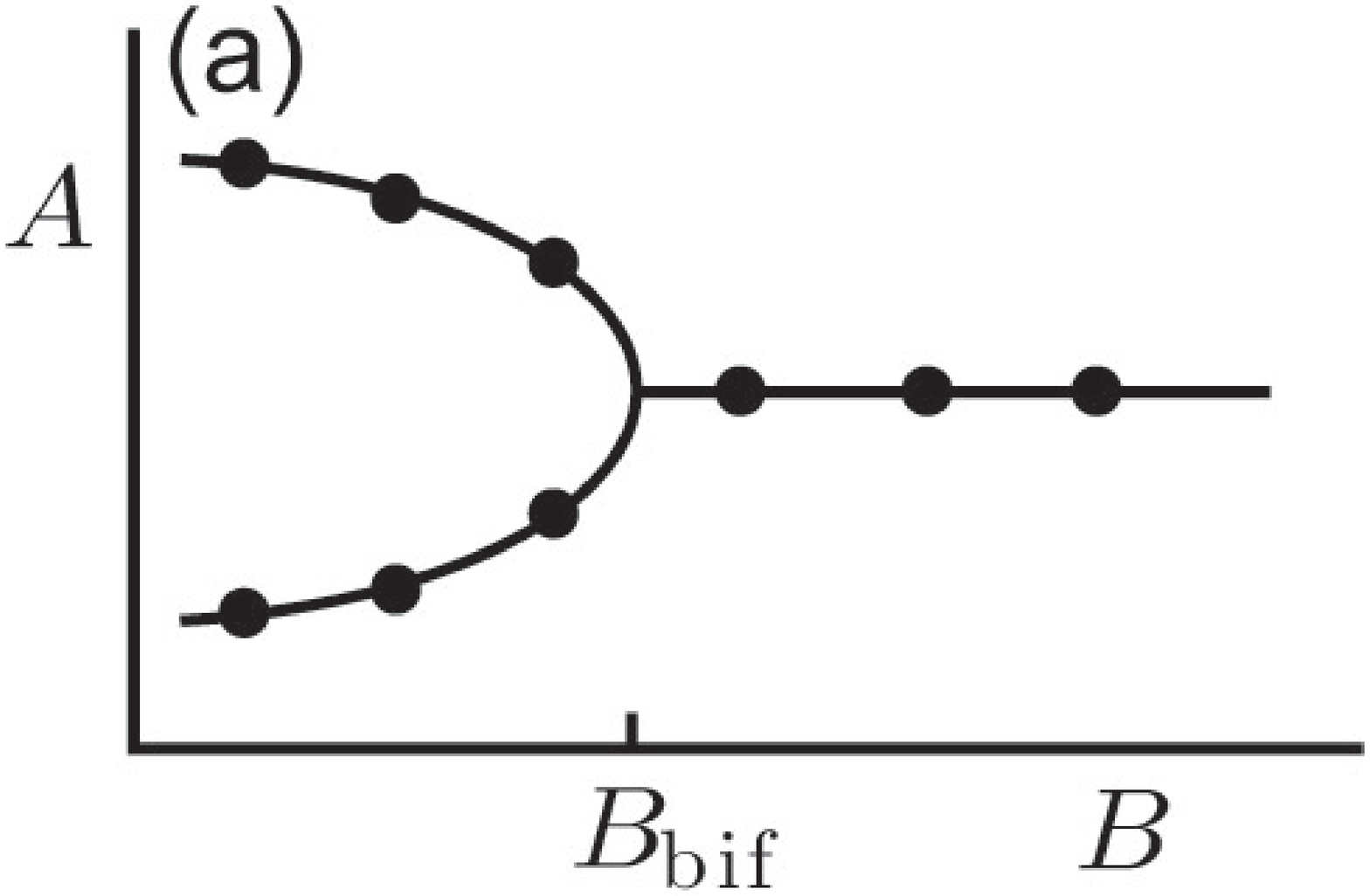} &
\includegraphics[width=1.5in]{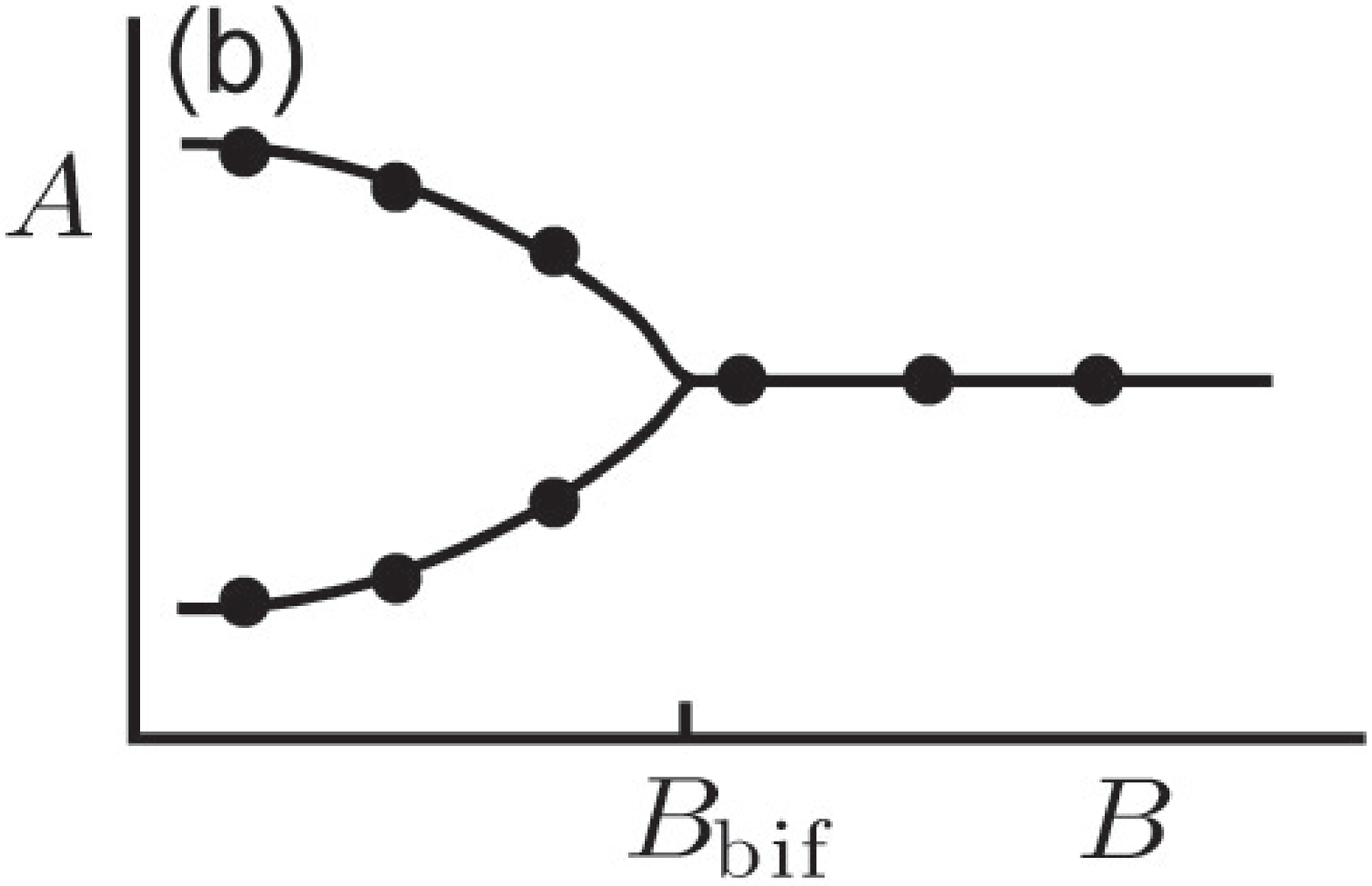}
\end{tabular}
\caption{Schematic bifurcation diagrams of period-doubling bifurcation: (a) a
smooth type and (b) a border-collision type. Here, $B$ represents a
bifurcation parameter and $A$ represents fixed-point solutions.}%
\label{fig:bifdiagrams}%
\end{figure}

The aforementioned intrinsic differences between the two bifurcation types
lead to differences in their bifurcation diagrams, as depicted in Fig.
\ref{fig:bifdiagrams}. Here, the two bifurcated branches of a smooth
period-doubling bifurcation become tangent to each other at the bifurcation
point, while the bifurcated branches of a border-collision bifurcation open at
an angle. Thus, in principle the bifurcation diagrams should distinguish
between the two bifurcation types. However, in practice, experiments can
provide only a limited number of measurements (especially in biological
systems), so the resulting bifurcation diagrams do not have sufficient
resolution. This is illustrated in Fig. \ref{fig:bifdiagrams}, where the
discrete points representing experimental data along a bifurcation diagram do
not readily reveal the true type of bifurcation. Therefore, there is a need
for a more sensitive technique to differentiate between the two bifurcations.

\subsection{Prebifurcation Amplification}

Based on prebifurcation amplification, our group has developed a robust
technique to distinguish between smooth and border-collision bifurcations
[27-30]. Here, we briefly review the results. It has been shown theoretically
and experimentally that, near the onset of a smooth period-doubling
bifurcation, subharmonic perturbations in a bifurcation parameter result in
amplified disturbances in the response, a phenomenon known as prebifurcation
amplification \cite{wiesenfeld85,wiesenfeld86,kravtsov03,surovyatkina04}. In
the following, we will show that, under variations in system parameters,
prebifurcation amplification exhibits qualitatively different scaling laws in
border-collision period-doubling and smooth period-doubling bifurcations.
Thus, prebifurcation amplification is a useful technique to distinguish
between the two possible types of period-doubling bifurcations.

To illustrate the concept of prebifurcation amplification, we consider a
dynamical system described by the following map
\begin{equation}
x_{n+1}=f\left(  x_{n};B\right)  , \label{eqn:generalMap}%
\end{equation}
where $B$ represents a bifurcation parameter, e.g., the BCL in
cardiac models. Both the function $f$ and state variable $x$ may be
one- or multi-dimensional. Let us assume that, at a critical value
$B=B_{\text{bif}}$, the system undergoes a period-doubling
bifurcation that is either a smooth type for smooth $f$
\cite{strogatz94,nayfeh95} or a border-collision type for piecewise
smooth $f$ \cite{diBernardo06,BCbook}. We further assume that the
stable period-one solution lies on the side $B>B_{\text{bif}}$, as
indicated in Fig. \ref{fig:bifdiagrams}.

When a subharmonic perturbation is applied to $B$ under conditions when
$B>B_{\text{bif}}$, it renders map (\ref{eqn:generalMap}) as%
\begin{equation}
x_{n+1}=f\left(  x_{n};B+\left(  -1\right)  ^{n}\delta\right)  ,
\label{eqn:app}%
\end{equation}
where $\delta$ is the amplitude of the perturbation. The perturbation may also
be imposed in the form of $B-\left(  -1\right)  ^{n}\delta$, which leads to a
solution only different in phase from that of Eqn. (\ref{eqn:app}). Since $B$
represents the pacing interval in cardiac models, such a variation in $B$ is
referred to as alternate pacing. For $B$ greater than but close to
$B_{\text{bif}}$ and small $\delta$, the steady-state response of Eqn.
(\ref{eqn:app}) consists of alternating recurrent states of $x_{\text{even}}$
and $x_{\text{odd}}$, which satisfy the following conditions%
\begin{align}
x_{\text{even}}  &  =f\left(  x_{\text{odd}};B-\delta\right)  ,\\
x_{\text{odd}}  &  =f\left(  x_{\text{even}};B+\delta\right)  .
\end{align}
In cardiac models, one component of the vector $x$ is the APD, henceforth
denoted by $A$. Alternate pacing of these models results in a long-short
beat-to-beat variation in pacing intervals, which in turns cause alternation
in $A$ even when $B>B_{\text{bif}}$. Since a period-doubling bifurcation is
sensitive to subharmonic perturbations, perturbations in $B$ result in
amplified disturbances in $A$. The effect of prebifurcation amplification can
then be characterized by a gain defined as follows
\begin{equation}
\text{$\Gamma$}\equiv\frac{\left\vert A_{\text{even}}-A_{\text{odd}%
}\right\vert }{2\delta}. \label{eqn:Gamma_def}%
\end{equation}

\subsubsection{Gain of Smooth Bifurcations}

Several authors \cite{heldstab83,wiesenfeld85,kravtsov03,surovyatkina04} have
investigated the influence of parameters on prebifurcation amplification in
smooth period-doubling bifurcations. In a previous paper \cite{zhao06}, we
explored the scaling laws between the amplification gain $\Gamma$ and the
parameters $B$ and $\delta$, using a mapping model of arbitrary dimension. It
was shown there that the gain of a smooth bifurcation satisfies the following
relation,%
\begin{equation}
c\,\delta^{2}\,\Gamma^{3}+\left(  B-B_{\text{bif}}\right)  \Gamma-\left\vert
k\right\vert =0, \label{eqn:gain_smooth}%
\end{equation}
where $c$ and $k$ are constants determined by the system's properties at the
bifurcation point. It was established that the gain is infinite if and only if
$B=B_{\text{bif}}$ and $\delta=0$. The rate of divergence as the parameters
tend to $\left(  B_{\text{bif}},0\right)  $ depends on the path taken. For
example, when $\delta$ is extremely small, the gain tends to infinity as
$\left(  B-B_{\text{bif}}\right)  ^{-1}$; on the other hand, when
$B=B_{\text{bif}}$, the gain tends to infinity as $\delta^{-2/3}$.

In cardiac experiments, it is very difficult to accurately locate the
bifurcation point. Moreover, the existence of noise and the limitation on the
number of measurements restrict one from using very small perturbations.
Instead, one can investigate the gains under two protocols: i) let $B$
approach $B_{\text{bif}}$ while retaining a finite and constant $\delta$; and
ii) let $\delta$ approach zero while retaining a constant $B>B_{\text{bif}}$.
As has been established in \cite{zhao06} that, under constant $\delta$,
$\Gamma$ scales according to $\left(  B-B_{\text{bif}}\right)  ^{-1}$ except
when $B-B_{\text{bif}}$ is sufficiently small, where the gain becomes
saturated. Alternatively, under constant $B>B_{\text{bif}}$, $\Gamma$ scales
to $\delta^{-2/3}$ except when $\delta$ is sufficiently small, where the gain
becomes saturated. Figure \ref{fig:gains} (a) and (b) schematically show the
behaviors of $\Gamma$ vs. $B$ and $\Gamma$ vs. $\delta$, respectively.

\subsubsection{Gain of Border-Collision Bifurcations}

The system (\ref{eqn:generalMap}) possesses a border-collision bifurcation if
the function $f$ is piecewise smooth as follows%
\begin{equation}
f\left(  x;B\right)  =\left\{
\begin{array}
[c]{cc}%
f_{1}\left(  x;B\right)  , & \text{if }h\left(  x\right)  <0\\
f_{2}\left(  x;B\right)  , & \text{if }h\left(  x\right)  >0
\end{array}
\right.  , \label{eqn:piecewisefcn}%
\end{equation}
where $h$ is a smooth scalar function and $h\left(  x\right)  =0$ indicates a
\textquotedblleft border\textquotedblright\ in the state space, on which
$f_{1}\left(  x;B\right)  =f_{2}\left(  x;B\right)  $. An approximate
expression for the gain is derived in the Appendix using a one-dimensional
map; results for general maps can be found in \cite{zhao06bc}. To lowest
order, the gain is piecewise smooth as follows
\begin{equation}
\text{$\Gamma$}=\left\{
\begin{array}
[c]{cc}%
\Gamma_{\text{const}}, & \text{if }\left(  B-B_{\text{bif}}\right)
/\delta>\rho_{\text{crit}}\\
\Gamma_{\text{const}}-\gamma\left(  \frac{B-B_{\text{bif}}}{\delta}%
-\rho_{\text{crit}}\right)  , & \text{if }\left(  B-B_{\text{bif}}\right)
/\delta<\rho_{\text{crit}}%
\end{array}
\right.  , \label{eqn:gain_bcb}%
\end{equation}
where $\Gamma_{\text{const}}$, $\gamma$, and $\rho_{\text{crit}}$ are positive
constants determined by system properties. Therefore, the gain is a constant
along any straight line $\left(  B-B_{\text{bif}}\right)  /\delta=const$.
Since all these lines intersect at $\left(  B_{\text{bif}},0\right)  $, the
gain at this point is not defined.

\begin{figure}[tbh]
\centering
\includegraphics[width=3in]{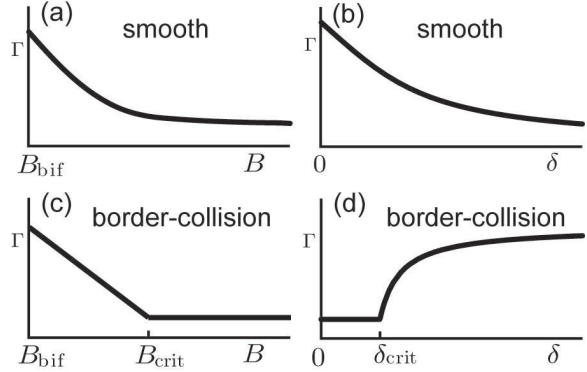}\caption{Prebifurcation gain $\Gamma$ of
a classical period-doubling bifurcation (a,b) and of a border-collision
period-doubling bifurcation (c,d). In panels (a) and (c), $\delta$ stays
constant; in panels (b) and (d), $B_{\text{bif}}<B$ stays constant. Comparison
between panels (b) and (d) provides the most revealing difference between the
two bifurcations.}%
\label{fig:gains}%
\end{figure}

Again, we apply the two protocols described in the previous subsection. When
$\delta$ is constant, the gain is constant when $B>B_{\text{crit}%
}=B_{\text{bif}}+\rho_{\text{crit}}\,\delta$ and varies linearly as $B$ when
$B<B_{\text{crit}}$. Alternatively, when $B$ is constant, the gain is constant
when $\delta<\delta_{\text{crit}}=\left(  B-B_{\text{bif}}\right)
/\rho_{\text{crit}}$ and varies as $\delta^{-1}$ when $\delta>\delta
_{\text{crit}}$. Schematic diagrams of these behaviors are shown in Figs.
\ref{fig:gains} (c) and (d). 

It is evident from Fig. \ref{fig:gains} that behaviors of the gain are
qualitatively different for a smooth bifurcation and for a border-collision
bifurcation. However, the differences between Figs. \ref{fig:gains} (a) and
(c) may be difficult to detect for discrete data or for data disturbed by
noise. Conversely, differences in Figs. \ref{fig:gains} (b) and (d) are
apparent even for discrete data and in the presence of noise. Therefore,
investigating the $\Gamma$ vs. $\delta$ under alternate pacing provides an
unambiguous way to distinguish between the two bifurcations. Moreover, since
this technique relies on the trend of the gain rather than the magnitude, it
allows one to distinguish between smooth and nonsmooth behaviors in
experiments without the need to accurately locate the bifurcation point
\cite{berger06}.

\subsection{Hybrid Behavior of the Prebifurcation Gain}

To identify the bifurcation mechanism mediating cardiac alternans,
we implemented the aforementioned technique in paced \emph{in vitro}
bullfrog heart \cite{Zhao2006ITP,berger06}, where the experiments
reveal a novel phenomenon that cannot be explained by the above
simple dichotomy of smooth/nonsmooth bifurcations. Specifically, our
experiments show that very close to the bifurcation point, $\Gamma$
decreases with $\delta$, which agrees with the smooth bifurcation
(Fig. 3b), whereas further away $\Gamma$ increases with $\delta$,
which agrees with the border-collision bifurcation (Fig. 3d). A
bifurcation that exibits such a crossover between smooth and
border-collision behaviors is named a hybrid period-doubling
bifurcation \cite{Zhao2006ITP,berger06}. We further found that the
essence of this hybrid behavior can be reproduced by a model of a
so-called unfolded border-collision bifurcation. In the remainder of
this paper, we will carry out a detailed analysis of the unfolded
border-collision model. This analysis will help to understand the
mathematical mechanism underlying the crossover between smooth and
nonsmooth behaviors, providing guidance for the development of
future models.

\section{A Model of an Unfolded Border-Collision Bifurcation}

We explore the mechanism of the aforementioned hybrid behavior using the
unfolded border-collision model presented in \cite{Zhao2006ITP,berger06}. For
illustration purpose, we first consider a piecewise smooth map%
\begin{equation}
A_{n+1}=A_{\text{c}}+\alpha\left(  D_{n}-D_{\text{th}}\right)  +\beta
\left\vert D_{n}-D_{\text{th}}\right\vert , \label{eqn:bc_map}%
\end{equation}
where $A_{n}$ and $D_{n}$ denote the $n$th action potential duration and
diastolic interval, respectively. Note that $D_{n}=B-A_{n}$ as can be seen
from Fig. \ref{fig:action_potential}. Under the following conditions (cf.
\cite{diBernardo06,BCbook})
\begin{equation}
-1<\alpha+\beta<1<\alpha-\beta\text{ and }-1<\alpha^{2}-\beta^{2}<1,
\label{eqn:bc_cond}%
\end{equation}
map (\ref{eqn:bc_map}) possesses a border-collision period-doubling
bifurcation at%
\begin{equation}
B_{c}=A_{\text{c}}+D_{\text{th}}.
\end{equation}
\ To remove the nonsmoothness of map (\ref{eqn:bc_map}), we \textquotedblleft
unfold\textquotedblright\ the singular term $\beta\left\vert D_{n}%
-D_{\text{th}}\right\vert $ as follows%

\begin{equation}
A_{n+1}=A_{\text{c}}+\alpha\left(  D_{n}-D_{\text{th}}\right)  +\beta
\sqrt{\left(  D_{n}-D_{\text{th}}\right)  ^{2}+D_{\text{s}}^{2}}.
\label{eqn:unfold}%
\end{equation}
Map (\ref{eqn:unfold}) represents a one-parameter family of maps that reduces
to map (\ref{eqn:bc_map}) when $D_{\text{s}}=0$. For any $D_{\text{s}}\neq0$,
the unfolded map (\ref{eqn:unfold}) is smooth and exhibits what is technically
a smooth period-doubling bifurcation. Nevertheless, the dynamics of map
(\ref{eqn:bc_map}) and map (\ref{eqn:unfold}) exhibit no significant
differences except when $B-B_{c}$ is less than or on the order of
$D_{\text{s}}$. It is worth noting that there are other ways to unfold the
border-collision map (\ref{eqn:bc_map}). Here, we choose map (\ref{eqn:unfold}%
) because of its simplicity and ease of analysis.

In the following, we show that map (\ref{eqn:unfold}) has a smooth
period-doubling bifurcation if $D_{\text{s}}\neq0$. To this end, we denote the
bifurcation point by $A=A_{\text{bif}}=B_{\text{bif}}-D_{\text{bif}}$ and let
the Jacobian of map (\ref{eqn:unfold}) equal to $-1$ at the bifurcation point;
in symbols%
\begin{equation}
-\alpha-\frac{\beta\left(  D_{\text{bif}}-D_{\text{th}}\right)  }%
{\sqrt{\left(  D_{\text{bif}}-D_{\text{th}}\right)  ^{2}+D_{\text{s}}^{2}}%
}=-1. \label{eqn:jac}%
\end{equation}
It follows from Eqn. (\ref{eqn:jac}) that%
\begin{equation}
\beta\left(  D_{\text{bif}}-D_{\text{th}}\right)  =\left(  1-\alpha\right)
\sqrt{\left(  D_{\text{bif}}-D_{\text{th}}\right)  ^{2}+D_{\text{s}}^{2}}.
\label{eqn:pdeq1}%
\end{equation}
Since $\beta<0$ as can be shown from the conditions in Eqn. (\ref{eqn:bc_cond}%
), the term $D_{\text{bif}}-D_{\text{th}}$ has an opposite sign as the term
$1-\alpha$; in symbols%
\begin{equation}
\left(  D_{\text{bif}}-D_{\text{th}}\right)  \left(  1-\alpha\right)  <0.
\label{eqn:Dbifsign}%
\end{equation}
Evaluating $D_{\text{bif}}$ from Eqn. (\ref{eqn:pdeq1}) and considering the
conditions (\ref{eqn:bc_cond}) and (\ref{eqn:Dbifsign}) yields
\begin{equation}
D_{\text{bif}}=D_{\text{th}}-\frac{\left(  1-\alpha\right)  D_{\text{s}}%
}{\sqrt{\beta^{2}-\left(  1-\alpha\right)  ^{2}}}.
\end{equation}
Thus, APD at the bifurcation point can be written as%
\begin{align}
A_{\text{bif}}  &  =A_{\text{c}}+\alpha\left(  D_{\text{bif}}-D_{\text{th}%
}\right)  +\beta\sqrt{\left(  D_{\text{bif}}-D_{\text{th}}\right)
^{2}+D_{\text{s}}^{2}},\nonumber\label{eqn:fixpt_bif}\\
&  =A_{\text{c}}+\alpha\left(  D_{\text{bif}}-D_{\text{th}}\right)
+\frac{\beta^{2}\left(  D_{\text{bif}}-D_{\text{th}}\right)  }{1-\alpha},\\
&  =A_{\text{c}}-\frac{\alpha\left(  1-\alpha\right)  +\beta^{2}}{\sqrt
{\beta^{2}-\left(  1-\alpha\right)  ^{2}}}D_{s},\nonumber
\end{align}
and the corresponding value of BCL is
\begin{align}
B_{\text{bif}}  &  \equiv A_{\text{bif}}+D_{\text{bif}},\nonumber\\
&  =A_{\text{c}}+D_{\text{th}}-\frac{\left(  1-\alpha^{2}+\beta^{2}\right)
D_{s}}{\sqrt{\beta^{2}-\left(  1-\alpha\right)  ^{2}}}.
\end{align}

\begin{figure}[tbh]
\centering
\includegraphics[width=2.5in]{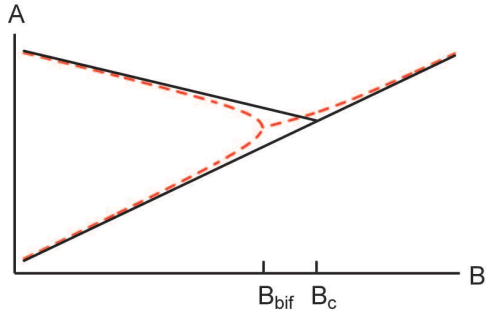}\caption{Schematic diagram showing a
border-collision bifurcation (solid) and the unfolded bifurcation (dashed).}%
\label{fig:unfold}%
\end{figure}

Comparing $B_{\text{bif}}$ and $B_{\text{c}}$ reveals that the smooth
period-doubling bifurcation in map (\ref{eqn:unfold}) reduces to the
border-collision period-doubling bifurcation in map (\ref{eqn:bc_map}) as
$D_{\text{s}}\rightarrow0$. Moreover, it can be shown from Eqn.
(\ref{eqn:bc_cond}) that $1-\alpha^{2}+\beta^{2}>0$ so that $B_{\text{bif}%
}<B_{\text{c}}$. Figure \ref{fig:unfold} demonstrates schematically the
relation between a border-collision bifurcation and the unfolded bifurcation.

\subsection{Analysis of the Response to Alternate Pacing}

To study the prebifurcation amplification of map (\ref{eqn:unfold}), we apply
an alternating perturbation to the BCL's; in symbol, $B_{n}=B+\left(
-1\right)  ^{n}\delta$, where $B$ is a baseline BCL and $\delta$ is a small
but nonzero perturbation. Under this alternate pacing, it follows that
$D_{n}=B+\left(  -1\right)  ^{n}\delta-A_{n}$. Here, we require that
$B>B_{\text{bif}}$ because prebifurcation dynamics is of interest. Denoting
the steady-state APDs under alternate pacing by $A_{\text{even}}$ and
$A_{\text{odd}}$, it follows from Eqn. (\ref{eqn:unfold}) that%
\begin{align}
A_{\text{even}}  &  =A_{\text{c}}+\alpha\left(  D_{\text{odd}}-D_{\text{th}%
}\right)  +\beta\sqrt{\left(  D_{\text{odd}}-D_{\text{th}}\right)
^{2}+D_{\text{s}}^{2}},\label{eqn:sseqn_even}\\
A_{\text{odd}}  &  =A_{\text{c}}+\alpha\left(  D_{\text{even}}-D_{\text{th}%
}\right)  +\beta\sqrt{\left(  D_{\text{even}}-D_{\text{th}}\right)
^{2}+D_{\text{s}}^{2}}, \label{eqn:sseqn_odd}%
\end{align}
where%
\begin{align}
D_{\text{odd}}  &  =B-\delta-A_{\text{odd}},\\
D_{\text{even}}  &  =B+\delta-A_{\text{even}}.
\end{align}
For later convenience, we let%
\begin{equation}
B=B_{c}+\Delta B=A_{\text{c}}+D_{\text{th}}+\Delta B
\end{equation}
and we define $\Delta_{\text{even}}$ and $\Delta_{\text{odd}}$ by%
\begin{equation}
A_{\text{even}}=\Delta_{\text{even}}+A_{\text{c}}\text{\thinspace\ \ and
}\ A_{\text{odd}}=\Delta_{\text{odd}}+A_{\text{c}}.
\end{equation}
Substituting the above equations into Eqns. (\ref{eqn:sseqn_even}) and
(\ref{eqn:sseqn_odd}) yields%
\begin{align}
\Delta_{\text{even}}+\alpha\,\left(  \Delta_{\text{odd}}+\delta-\Delta
B\right)   &  =\beta\sqrt{\left(  \Delta_{\text{odd}}+\delta-\Delta B\right)
^{2}+D_{\text{s}}^{2}},\\
\Delta_{\text{odd}}+\alpha\,\left(  \Delta_{\text{even}}-\delta-\Delta
B\right)   &  =\beta\sqrt{\left(  \Delta_{\text{even}}-\delta-\Delta B\right)
^{2}+D_{\text{s}}^{2}}.
\end{align}
One can then show that%
\begin{align}
&  \left(  \left(  1-\alpha\right)  \left(  \Delta_{\text{even}}%
-\Delta_{\text{odd}}\right)  +2\,\alpha\,\delta\right)  \left(  \left(
1+\alpha\right)  \left(  \Delta_{\text{even}}-\Delta_{\text{odd}}\right)
-2\,\alpha\,\Delta B\right) \nonumber\\
&  =\beta^{2}\left(  \Delta_{\text{odd}}-\Delta_{\text{even}}+2\,\delta
\right)  \left(  \Delta_{\text{odd}}+\Delta_{\text{even}}-2\,\Delta B\right)
\label{eqn:relation}%
\end{align}
Let
\begin{equation}
\gamma=\frac{\Delta_{\text{even}}-\Delta_{\text{odd}}}{2\,\delta};
\end{equation}
i.e., $\gamma$ is a gain-like quantity that can be either positive or negative
(cf. \ref{eqn:Gamma_def}). Substituting the definition of $\gamma$ into Eqn.
(\ref{eqn:relation}) yields%
\begin{align}
&  2\,\delta\left(  \left(  1-\alpha\right)  \gamma+\alpha\right)  \left(
\left(  1+\alpha\right)  \left(  \Delta_{\text{even}}-\Delta_{\text{odd}%
}\right)  -2\,\alpha\,\Delta B\right) \nonumber\\
&  =2\delta\,\beta^{2}\left(  1-\gamma\right)  \left(  \Delta_{\text{odd}%
}+\Delta_{\text{even}}-2\,\Delta B\right)  . \label{eqn:2delta}%
\end{align}
Because we consider nonzero $\delta$, Eqn. (\ref{eqn:2delta}) can be reduced
to%
\begin{equation}
\gamma\left(  c_{1}\Delta B+d_{1}\left(  \Delta_{\text{even}}+\Delta
_{\text{odd}}\right)  \right)  =c_{2}\Delta B+d_{2}\left(  \Delta
_{\text{even}}+\Delta_{\text{odd}}\right)  ,
\end{equation}
where%
\begin{align}
c_{1}  &  =-2\left(  \beta^{2}+\alpha\left(  1-\alpha\right)  \right)  ,\\
d_{1}  &  =1-\alpha^{2}+\beta^{2},\\
c_{2}  &  =2\left(  \alpha^{2}-\beta^{2}\right)  ,\\
d_{2}  &  =-\left(  \alpha\left(  1+\alpha\right)  -\beta^{2}\right)  .
\end{align}
It then follows that%
\begin{equation}
\gamma=\frac{c_{2}\Delta B+d_{2}\left(  \Delta_{\text{even}}+\Delta
_{\text{odd}}\right)  }{c_{1}\Delta B+d_{1}\left(  \Delta_{\text{even}}%
+\Delta_{\text{odd}}\right)  }.
\end{equation}
Since $\Delta_{\text{even}}$ and $\Delta_{\text{odd}}$ depend on $\Delta B$
and $\delta$, $\gamma$ is a function of $\Delta B$ and $\delta$.

Recalling definition (\ref{eqn:Gamma_def}), we find the prebifurcation
amplification gain as%
\begin{equation}
\Gamma=\left\vert \frac{c_{2}\Delta B+d_{2}\left(  \Delta_{\text{even}}%
+\Delta_{\text{odd}}\right)  }{c_{1}\Delta B+d_{1}\left(  \Delta_{\text{even}%
}+\Delta_{\text{odd}}\right)  }\right\vert .
\end{equation}
Particularly, when $\Delta B=0$, i.e. $B=B_{\text{c}}=A_{\text{c}%
}+D_{\text{th}}$, the gain is%
\begin{equation}
\Gamma=\left\vert \frac{d_{2}}{d_{1}}\right\vert =\frac{\alpha+\left(
\alpha^{2}-\beta^{2}\right)  }{1-\left(  \alpha^{2}-\beta^{2}\right)  }.
\end{equation}
Therefore, when $B=B_{c}$, $\Gamma$ is same for all $\delta$. With some
manipulation, one can show that $\partial\Gamma/\partial B\neq0$ at $B=B_{c}$.
Moreover, when $B$ is sufficiently close to $B_{\text{bif}}<B_{c}$ and
$\delta$ is fixed, $\Gamma$ decreases as $B$ increases as described in
previous section and proven in \cite{zhao06}. Thus, for a given $\delta$,
$\Gamma$ is a monotonically decreasing function of $B$ and $\Gamma$ becomes
constant at $B=B_{c}$. Because $\Gamma$ is a monotonically decreasing function
of $\delta$ when $B\gtrsim B_{\text{bif}}$, as shown in the previous section
(see also \cite{zhao06}), it follows by continuity that $\Gamma$ will increase
as $\delta$ increases in the region of $B>B_{c}$. In other words, the map
(\ref{eqn:unfold}) exhibits smooth like behavior when $B $ is sufficiently
close to $B_{\text{bif}}$ and border-collision like behavior when $B>B_{c}$
(see the relation between $B_{\text{bif}}$ and $B_{\text{c}}$ in Fig.
\ref{fig:unfold}).

\subsection{Numerical Example}

\begin{figure}[tbh]
\centering
\begin{tabular}
[c]{c}%
\includegraphics[width=2.5in]{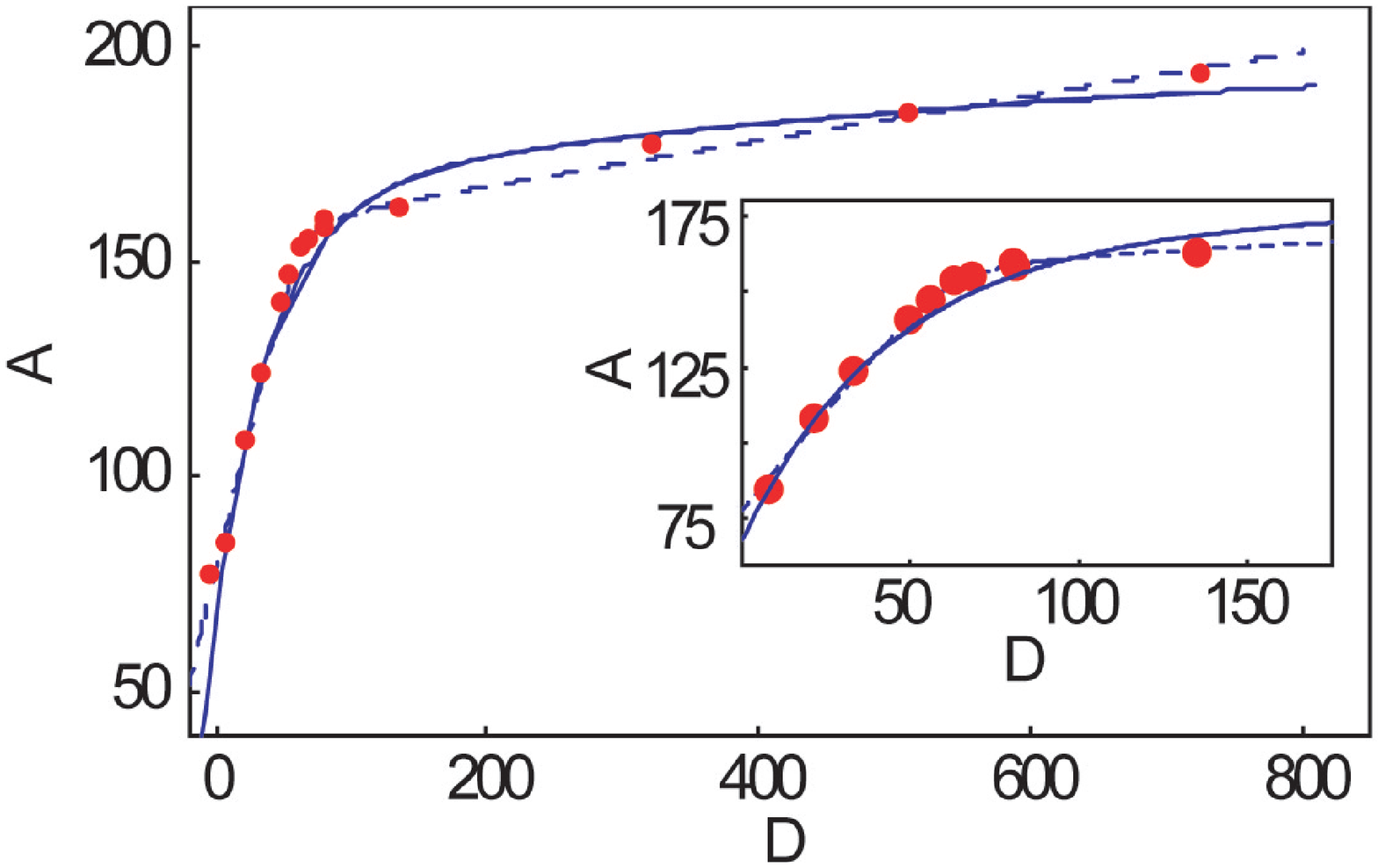}\\
\includegraphics[width=2.5in]{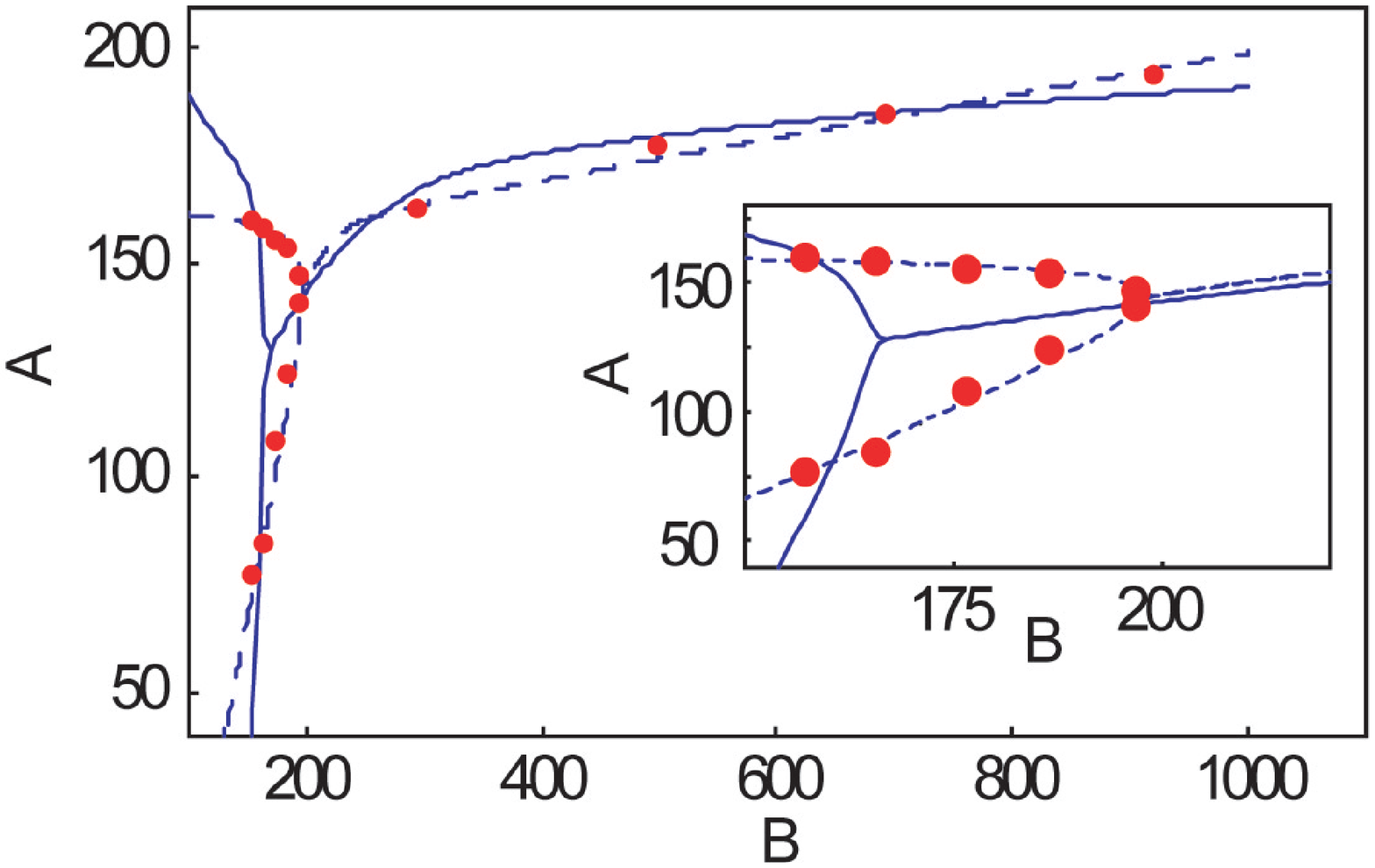}
\end{tabular}
\caption{Comparison between the model of Guevara \emph{et al.} (solid) and the
unfolded border-collision model (dashed) in fitting the experimental data in
\cite{Guevara84} (points). Although both models fit the dynamic restitution
curve well (top panel), the unfolded border-collision model fits alternans
data much better (bottom panel).}%
\label{fig:fit}%
\end{figure}

Before comparing the proposed model to experimental data, we review the class
of models that are most commonly used in the cardiac research community. These
models relate APD and DI through exponential functions. Typically, parameters
of a model are obtained by fitting the model to the so-called dynamic
restitution curve, which is a plot of the steady-state APD vs. DI. For
example, in their pioneering work, Guevara \emph{et al.} \cite{Guevara84}
proposed a model of cardiac dynamics as%
\begin{equation}
A_{n+1}=201-98\,e^{-D_{n}/43}-35\,e^{-D_{n}/653}, \label{eqn:model_gue}%
\end{equation}
where all variables and parameters have the unit of millisecond. Parameters of
map (\ref{eqn:model_gue}) were obtained by fitting the dynamic restitution
curve measured in experiments performed on quiescent aggregates of ventricular
cells from 7-day-old embryonic chick hearts \cite{Guevara84}. Although the
model of Guevara \emph{et al.} fits the dynamic restitution curve reasonably
well (Fig. \ref{fig:fit}, top panel), it does not accurately describe the
response beyond the bifurcation to alternans, as is evident from the
bifurcation diagram of steady-state APD vs. BCL (Fig. \ref{fig:fit}, bottom
panel). A careful examination reveals that the dynamic restitution curve is
well approximated by two distinct parts with significantly different slopes.
The transition between the two slopes occurs with a small interval near
DI$\approx60$ ms, which is also approximately where the transition to
alternans occurs. Now, we recall that the unfolded border-collision model
(\ref{eqn:unfold}), with properly chosen parameters, describes such rapid
changes between two distinct slopes. Fitting map (\ref{eqn:unfold}) to the
experimental dynamic restitution data, we obtain a set of parameters%
\begin{gather}
\alpha=0.69,\,\beta=-0.64,\,\nonumber\\
A_{\text{c}}=161\,\text{ms},\,D_{\text{th}}=62\,\text{ms,}\,\text{and}%
\ D_{\text{s}}=15\,\text{ms}.
\end{gather}
As shown in Fig. \ref{fig:fit}, the unfolded border-collision map
(\ref{eqn:unfold}) with these parameters faithfully reproduces the bifurcation
diagram, including the alternans branches. As demonstrated in Fig.
\ref{fig:unfold}, the bifurcation diagram of a border-collision map is close
to that of its unfolded counterpart except near the bifurcation point. Thus,
one expects a reasonable fit to the experimental data in Fig. \ref{fig:fit}
using a pure border-collision map, i.e., letting $D_{s}=0$. However, as has
been established in the previous section, the nonsmooth map can not capture
hybrid behaviors in the prebifurcation gain. Here, for clarity, the
bifurcation diagram of the corresponding border-collision model is not shown
in Fig. \ref{fig:fit}.

\begin{figure}[tbh]
\centering
\begin{tabular}
[c]{c}%
\includegraphics[width=2.5in]{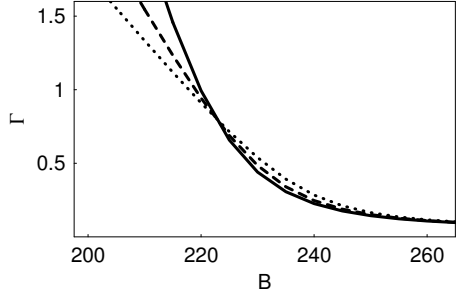}
\end{tabular}
\caption{Prebifurcation amplification predicted by the unfolded
border-collision model (\ref{eqn:unfold}): $\delta=5$ ms (solid), $\delta=10$
ms (dashed), and $\delta=15$ ms (dotted).}%
\label{fig:modelgain}%
\end{figure}

We then simulate map (\ref{eqn:unfold}) with alternate pacing. Figure
\ref{fig:modelgain} shows $\Gamma$ vs. $B$ for different values of $\delta$.
These curves cross one another at $B=B_{\text{c}}=223$ ms. Note that a
period-doubling bifurcation occurs at $B=198$ ms. It is clear that $\Gamma$
vs. $\delta$ displays a trend consistent with a smooth bifurcation (cf. Fig.
\ref{fig:gains} (b)) when $B<B_{\text{c}}$ and, on the other hand, $\Gamma$
vs. $\delta$ shows a trend consistent with a border-collision bifurcation when
$B>B_{\text{c}}$ (cf. Fig. \ref{fig:gains} (d)). Since Guevara \emph{et al.}
did not perform alternate pacing experiments, no data are available for
comparison. However, we note that the simulation here is in qualitative
agreement with our previous experiments on bullfrog ventricles
\cite{Zhao2006ITP,berger06}. 

\section{Discussion and Conclusion}

Theoretical analysis of the prebifurcation amplification reveals that
different scaling laws are associated with smooth and border-collision
period-doubling bifurcations. The differences appear in the following three
aspects. First, the gain of a smooth bifurcation tends to infinity as $\left(
B,\delta\right)  $ approaches $\left(  B_{\text{bif}},0\right)  $; conversely,
the gain of a border-collision bifurcation is finite everywhere but not
defined at $\left(  B_{\text{bif}},0\right)  $. Second, the gain of a smooth
bifurcation varies smoothly under changes in system parameters while that of a
border-collision bifurcation undergoes a nonsmooth variation as parameters
cross a boundary in the parameter space (see Figs. \ref{fig:gains} (a,b) and
\ref{fig:gains} (c,d)). Third, under constant $B$ and increasing $\delta$, the
gain of a smooth bifurcation decreases while that of a border-collision
bifurcation increases. Thus, the gain versus perturbation size relation
provides a more sensitive criterion to differentiate between the two
bifurcation types. As can be seen from Fig. \ref{fig:gains} (b) and (d), even
with few data points, the $\Gamma$ vs. $\delta$ relation clearly reveals the
underlying bifurcation mechanism. On the other hand, the bifurcation diagram
does not allow one to distinguish between the two bifurcations with only a few
data points nor does the $\Gamma$ vs. $B$ relation.

Although the technique described here was developed with a goal of
identifying the type of bifurcation mediating alternans, the
analysis is based on general iterated maps. Thus, the results are
independent of any physical details of cardiac dynamics and can be
readily applied to any dynamical systems.

The analysis based on simple dichotomy of smooth/border-collision bifurcation
has limitations. Since it is assumed that a system either has well behaved
derivatives or is discontinuous in first derivatives, the result is not
directly applicable to the intermediate case, i.e., a system whose first
derivatives are continuous but change rapidly. The model of unfolded
border-collision bifurcation studied here serves to address the latter case.

Previous experimental findings \cite{Zhao2006ITP,berger06} suggest that
modeling of cardiac dynamics should consider the rapid changes in the system's
properties, i.e., large variations over a narrow parameter interval. As one
example, we study here the smoothed version of a border-collision model. We
show that the smoothed map indeed unfolds the original border-collision
period-doubling bifurcation to a smooth one. In addition, we carry out the
analysis of the unfolded map under alternate pacing. The result indicates that
the unfolded border-collision model exhibits hybrid smooth/nonsmooth
behaviors, which is in qualitative agreement with previous experimental
observations on bullfrog hearts \cite{Zhao2006ITP,berger06}. We further
illustrate that the unfolded border-collision model can more accurately
describe alternans observed in an experiment on embryonic chick hearts
\cite{Guevara84}. The fact that hybrid behaviors are observed in different
species indicates that this phenomenon may be prevalent in cardiac dynamics.
It is worth noting that, besides the model studied here, the crossover between
smooth and nonsmooth behaviors can also be captured by other types of maps. We
choose the current model solely based on its simplicity and ease of analysis.

We note that many other physical systems also possess rapid changes in
systems' properties. To fully describe such rapid changes, one would need to
use functions with highly localized properties. For convenience of analysis,
these highly localized functions are often replaced by piecewise smooth
functions, where each piece adopts a much simpler form. Perhaps, the simplest
example is a bouncing ball, whose velocity changes rapidly before and after
impacts and is often modeled by an instantaneous jump using the coefficient of
restitution (see other examples in a recent special issue of the journal
Nonlinear Dynamics on discontinuous dynamical systems \cite{Shukla2007}).
Although this approach has proven to be useful in many problems in engineering
and science, it brings up a more subtle question on the relation between the
piecewise smooth bifurcation problem and the original smooth bifurcation
problem. In \cite{Dankowicz2007ND}, Dankowicz purposefully coarsened a smooth
vector field with a piecewise smooth one and compared their bifurcation
diagrams. The full potential and limitations of the idea of intentional
nonsmoothing of a smooth function need to be explored in future research.

\begin{center}
\textbf{Acknowledgments}
\end{center}

Support of the National Institutes of Health under grant 1R01-HL-72831 and the
National Science Foundation under grants DMS-9983320 and PHY-0549259 is
gratefully acknowledged.

\section*{Appendix: Alternate Pacing of a Border-Collision Map}

In a previous paper \cite{zhao06bc}, we have shown the general results of
prebifurcation amplification for border-collision bifurcations using
high-dimensional maps. Here, we briefly review the results using a
one-dimensional map for simplicity. Consider a one-dimensional piecewise
continuous map of $A$ with a bifurcation parameter $B$ as follows%

\begin{equation}
A_{n+1}=\left\{
\begin{array}
[c]{cc}%
f_{1}\left(  A_{n};B\right)  , & \text{if }A_{n}>A_{\text{bif}}\\
f_{2}\left(  A_{n};B\right)  , & \text{if }A<A_{\text{bif}}%
\end{array}
\right.  , \label{eqn:pwcmap}%
\end{equation}
where $f_{1}\left(  A;B\right)  =f_{2}\left(  A;B\right)  $ when
$A=A_{\text{bif}}$. Assume a border-collision bifurcation occurs at
$B=B_{\text{bif}}$ and $A=A_{\text{bif}}$, as indicated in Fig.
\ref{fig:bifdiagrams} (b). Then the following conditions are satisfied at the
bifurcation point%
\begin{align}
\partial_{A}f_{2\,}  &  <-1<\partial_{A}f_{1\,}<1,\label{eqn:condition1}\\
0  &  <\partial_{B}f_{1}=\partial_{B}f_{2}\equiv\partial_{B}f,
\label{eqn:condition2}%
\end{align}
where all derivatives are evaluated at the bifurcation point $\left(
A_{\text{bif}};B_{\text{bif}}\right)  $. For conditions on border-collision
period-doubling bifurcations in multi-dimensional maps, see
\cite{diBernardo06,zhao06bc}.

Alternate pacing changes the map (\ref{eqn:pwcmap}) to%
\begin{equation}
A_{n+1}=\left\{
\begin{array}
[c]{cc}%
f_{1}\left(  A_{n};B+\left(  -1\right)  ^{n}\delta\right)  , & \text{if }%
A_{n}>A_{\text{bif}}\\
f_{2}\left(  A_{n};B+\left(  -1\right)  ^{n}\delta\right)  , & \text{if }%
A_{n}<A_{\text{bif}}%
\end{array}
\right.  . \label{eqn:pwsapp}%
\end{equation}
Due to the alternating perturbation, the steady state of Eqn.
(\ref{eqn:pwsapp}) is a period-two solution, whose two branches can be written
as
\begin{equation}
A_{n}=\left\{
\begin{array}
[c]{cc}%
A_{\text{odd}}\left(  B,\delta\right)  , & \text{for odd }n\\
A_{\text{even}}\left(  B,\delta\right)  , & \text{for even }n
\end{array}
\right.  .
\end{equation}
Particularly,
\begin{equation}
A_{\text{odd}}\left(  B_{\text{bif}},0\right)  =A_{\text{even}}\left(
B_{\text{bif}},0\right)  =A_{\text{bif}}.
\end{equation}
This solution consists of two different types: 1) in a unilateral solution,
both branches are above the border, i.e., $A_{\text{even}}>A_{\text{bif}}$ and
$A_{\text{odd}}>A_{\text{bif}}$; 2) in a bilateral solution, one branch is
above and the other branch below the border, i.e., $\left(  A_{\text{even}%
}-A_{\text{bif}}\right)  \times\left(  A_{\text{odd}}-A_{\text{bif}}\right)
<0$. In the following, we restrict attention to $B\geq B_{\text{bif}}$
(prebifurcation condition) and deal with the two types of solutions, respectively.

\subsection*{Unilateral Solution}

Because $A_{\text{even}}>A_{\text{bif}}$ and $A_{\text{odd}}>A_{\text{bif}}$,
it follows from Eqn. (\ref{eqn:pwsapp}) that%
\begin{align}
A_{\text{even}}  &  =f_{1}\left(  A_{\text{odd}};B-\delta\right)
,\label{eqn:md_border_f11}\\
A_{\text{odd}}  &  =f_{1}\left(  A_{\text{even}};B+\delta\right)  .
\label{eqn:md_border_f12}%
\end{align}
To leading order, the solution of Eqs. (\ref{eqn:md_border_f11}) and
(\ref{eqn:md_border_f12}) is
\begin{align}
A_{\text{even}}  &  =A_{\text{bif}}+\frac{\partial_{B}f}{1-\partial_{A}f_{1}%
}\left(  B-B_{\text{bif}}\right)  -\frac{\partial_{B}f}{1+\partial_{A}f_{1}%
}\delta,\label{eqn:xeven_caseA}\\
A_{\text{odd}}  &  =A_{\text{bif}}+\frac{\partial_{B}f}{1-\partial_{A}f_{1}%
}\left(  B-B_{\text{bif}}\right)  +\frac{\partial_{B}f}{1+\partial_{A}f_{1}%
}\delta. \label{eqn:xodd_caseA}%
\end{align}
Recalling the conditions in Eqs. (\ref{eqn:condition1}) and
(\ref{eqn:condition2}), it follows that $A_{\text{odd}}>A_{\text{even}}$.
Moreover, it follows from Eqn. (\ref{eqn:xeven_caseA}) that the unilateral
solution is valid as long as
\begin{equation}
B-B_{\text{bif}}>\rho_{\text{crit}}\,\delta,
\end{equation}
where%
\begin{equation}
\rho_{\text{crit}}=\frac{1-\partial_{A}f_{1}}{1+\partial_{A}f_{1}}>0\text{.}%
\end{equation}

\subsection*{Bilateral Solution}

The bilateral solution occurs in the region $B-B_{\text{bif}}<\rho
_{\text{crit}}\,\delta$. By continuity, the solution in this region satisfies
$A_{\text{odd}}>A_{\text{bif}}>A_{\text{even}}$. It follows from Eqn.
(\ref{eqn:pwsapp}) that%
\begin{align}
A_{\text{even}}  &  =f_{1}\left(  A_{\text{odd}};B-\delta\right)
,\label{eqn:md_border_f}\\
A_{\text{odd}}  &  =f_{2}\left(  A_{\text{even}};B+\delta\right)  .
\label{eqn:md_border_g}%
\end{align}
Linearizing Eqns. (\ref{eqn:md_border_f}) and (\ref{eqn:md_border_g}) around
$A=A_{\text{bif}}$ and $B=B_{\text{bif}}$ yields%
\begin{align}
A_{\text{even}}  &  =A_{\text{bif}}+\partial_{A}f_{1}\ast\left(
A_{\text{odd}}-A_{\text{bif}}\right)  +\partial_{B}f\ast\left(
B-B_{\text{bif}}-\delta\right)  ,\\
A_{\text{odd}}  &  =A_{\text{bif}}+\partial_{A}f_{2}\ast\left(  A_{\text{even}%
}-A_{\text{bif}}\right)  +\partial_{B}f\ast\left(  B-B_{\text{bif}}%
+\delta\right)  ,
\end{align}
where the derivatives are evaluated at $\left(  A_{\text{bif}};B_{\text{bif}%
}\right)  $. Solving the above equations yields the leading-order solution for
$A_{\text{even}}$ and $A_{\text{odd}}$ as%
\begin{align}
A_{\text{even}}  &  =A_{\text{bif}}+\frac{\left(  1+\partial_{A}f_{1}\right)
\left(  B-B_{\text{bif}}\right)  -\,\left(  1-\partial_{A}f_{1}\right)
\delta}{1-\partial_{A}f_{2}\,\partial_{A}f_{1}}\partial_{B}%
f,\label{eqn:xeven_caseB}\\
A_{\text{odd}}  &  =A_{\text{bif}}+\frac{\left(  1+\partial_{A}f_{2}\right)
\left(  B-B_{\text{bif}}\right)  +\left(  1-\partial_{A}f_{2}\right)  \delta
}{1-\partial_{A}f_{2}\,\partial_{A}f_{1}}\partial_{B}f. \label{eqn:xodd_caseB}%
\end{align}

\subsection*{Prebifurcation Gain}

When $B-B_{\text{bif}}>\rho_{\text{crit}}\,\delta$, it follows from Eqs.
(\ref{eqn:xeven_caseA}) and (\ref{eqn:xodd_caseA}) that the gain is%
\begin{equation}
\Gamma=\frac{A_{\text{odd}}-A_{\text{even}}}{2\,\delta}=\frac{\partial_{B}%
f}{1+\partial_{A}f_{1}}\equiv\Gamma_{\text{const}}.
\end{equation}
When $B-B_{\text{bif}}<\rho_{\text{crit}}\,\delta$, it follows from Eqs.
(\ref{eqn:xeven_caseB}) and (\ref{eqn:xodd_caseB}) that the gain is%
\begin{align}
\Gamma &  =\frac{A_{\text{odd}}-A_{\text{even}}}{2\,\delta}\\
&  =\Gamma_{\text{const}}-\gamma\left(  \frac{B-B_{\text{bif}}}{\delta}%
-\rho_{\text{crit}}\right)  ,
\end{align}
where%
\begin{equation}
\gamma=\frac{\partial_{A}f_{1}-\partial_{A}f_{2}}{2\left(  1-\partial_{A}%
f_{2}\,\partial_{A}f_{1}\right)  }\partial_{B}f>0.
\end{equation}

\end{document}